\documentclass[prl,twocolumn,superscriptaddress,floatfix,noshowpacs,10pt,longbibliography]{revtex4-2}
\usepackage{graphicx,bm,times}
\usepackage{amsmath}
\usepackage{amsfonts}
\usepackage{amssymb}
\usepackage{bm}% bold math
\usepackage{color}
\usepackage{times}
\usepackage[%
  colorlinks=true,
  urlcolor=blue,
  linkcolor=blue,
  citecolor=blue
  ]{hyperref}
\DeclareUnicodeCharacter{2212}{-}
\DeclareUnicodeCharacter{2009}{\,}

\begin{document}

\title{The drastic effect of the impurity scattering on the electronic and \\ superconducting properties of Cu-doped FeSe }

\author{Z. Zajicek}
\email[corresponding author:]{zachary.zajicek@physics.ox.ac.uk}
\affiliation{Clarendon Laboratory, Department of Physics, University of Oxford, Parks Road, Oxford OX1 3PU, UK}

\author{S. J. Singh}
\affiliation{Clarendon Laboratory, Department of Physics,
University of Oxford, Parks Road, Oxford OX1 3PU, UK}

\author{H. Jones}
\affiliation{Clarendon Laboratory, Department of Physics,
University of Oxford, Parks Road, Oxford OX1 3PU, UK}

\author{P. Reiss}
\affiliation{Clarendon Laboratory, Department of Physics,
University of Oxford, Parks Road, Oxford OX1 3PU, UK}
\affiliation{Max-Planck Institute for Solid State Research, Heisenbergstraße 1, 70569 Stuttgart, Germany}

\author{M. Bristow}
\affiliation{Clarendon Laboratory, Department of Physics, University of Oxford, Parks Road, Oxford OX1 3PU, UK}

\author{A. Martin}
\affiliation{Clarendon Laboratory, Department of Physics, University of Oxford, Parks Road, Oxford OX1 3PU, UK}

\author{A. Gower}
\affiliation{Clarendon Laboratory, Department of Physics, University of Oxford, Parks Road, Oxford OX1 3PU, UK}

\author{A. McCollam}
\affiliation{High Field Magnet Laboratory (HFML-EMFL), Radboud University, 6525 ED Nijmegen, The Netherlands}

\author{A. I. Coldea}
\email[corresponding author:]{amalia.coldea@physics.ox.ac.uk}
\affiliation{Clarendon Laboratory, Department of Physics, University of Oxford, Parks Road, Oxford OX1 3PU, UK}

\begin{abstract}
Non-magnetic impurities in iron-based superconductors can provide an important tool to understand the
pair symmetry and  they can influence significantly the transport and the superconducting behaviour.
Here, we present a study  of the role of strong impurity potential in the Fe plane,
induced by Cu substitution, on the electronic and superconducting properties of single crystals of FeSe.
The addition of Cu quickly suppresses both the nematic and superconducting states, and increases the residual resistivity due to enhanced impurity scattering.
Using magnetotransport data up to 35~T for a small amount of Cu impurity, we detect a significant reduction in the mobility of the charge carriers by a factor of $\sim 3$.
While the electronic conduction is strongly disrupted by Cu substitution, we identify additional signatures of anisotropic scattering which manifest in linear resistivity at low temperatures and $H^{1.6}$ dependence of magnetoresistance.
The suppression of superconductivity by Cu substitution is consistent with a sign-changing $s_{\pm}$ order parameter.
Additionally, in the presence of compressive strain, the superconductivity is enhanced, similar to FeSe.

\end{abstract}
\date{\today}
\maketitle

\section{Introduction}
Substitution of transition metals in different iron-based superconductors, like the parent compound BaFe$_{2}$As$_{2}$,
reveals the presence of the structural and magnetic transitions which are suppressed
while a robust superconducting dome is stabilized.
 Whilst many transition metals, like Co or Ni stabilise superconductivity \cite{Canfield2010,Saha2009},
the Cu substitution displays an unusual behaviour from being a dopant of electrons or holes
\cite{Canfield2010,Cui2013,Skornyakov2021}, causing major changes in transport behaviour and local magnetism \cite{Pelliciari2021}.
In NaFeAs, Cu substitution leads to an increase in resistivity, and a metal-to-insulator transition \cite{Pelliciari2021,Wang2013}
which leads to a decrease in  the spectral weight \cite{Cui2013}.
This type of transition was linked to a Mott-like insulator phase suggesting that Cu substitution
can be a tuning parameter towards strong correlations \cite{Pelliciari2021,Skornyakov2021}.

Among iron-based superconductors, FeSe displays
 a nematic electronic phase  below 90~K
but no long-range spin-density wave is stabilised at ambient pressure \cite{Coldea2017}.
 Despite the lack of long-range magnetism, a large range of spin excitations are present \cite{Kreisel2020}. The low-energy spin fluctuations 
can stabilize the anisotropic superconductivity \cite{Sprau2017} of FeSe
 and can affect the low temperature normal transport properties \cite{Bristow2020}.
By using isovalent substitution via S or Te for Se outside the conducting Fe plane
the nematic electronic phase is suppressed, whilst the superconductivity
remains rather robust over a large compositional range \cite{Coldea2021}.
In contrast to other iron-based superconductors,
the substitution with all transition metals inside the Fe plane
 leads to the suppression of both nematicity and superconductivity
 \cite{Mizuguchi2008,Urata2016,Williams2009,Huang2010,Gong2021}.
Additionally, the Cu-substitution in FeSe
leads to a metal-to-insulator transition for small substitutions ($x = 0.04$)
 \cite{Huang2010,Williams2009,Young2010}
 and it can induce local magnetism around the Cu sites for higher substitutions \cite{Williams2009}.
Under high pressure, the insulating behaviour is suppressed and the superconductivity is restored
as the magnetic fraction
 \cite{Schoop2011,Shylin2018,Deng2021}.
 Thus, the Cu substitution in FeSe
can reveal important information about the nature of the superconducting and normal states,
and whether an insulating state can be tuned into a high-$T_{\rm c}$ superconductor
under applied pressure in an iron-based superconductor.
Furthermore, the strong impurity scattering effects
induced by the substitution of various transition metal ions in the Fe plane can be compared with those
induced by irradiation, where superconductivity of FeSe was found to be enhanced by defects \cite{Teknowijoyo2016},
thus raising further questions about its pairing symmetry in the presence of disorder.

\begin{figure*}[htbp]
	\centering
	\includegraphics[trim={0cm 0cm 0cm 0cm}, width=1\linewidth,clip=true]{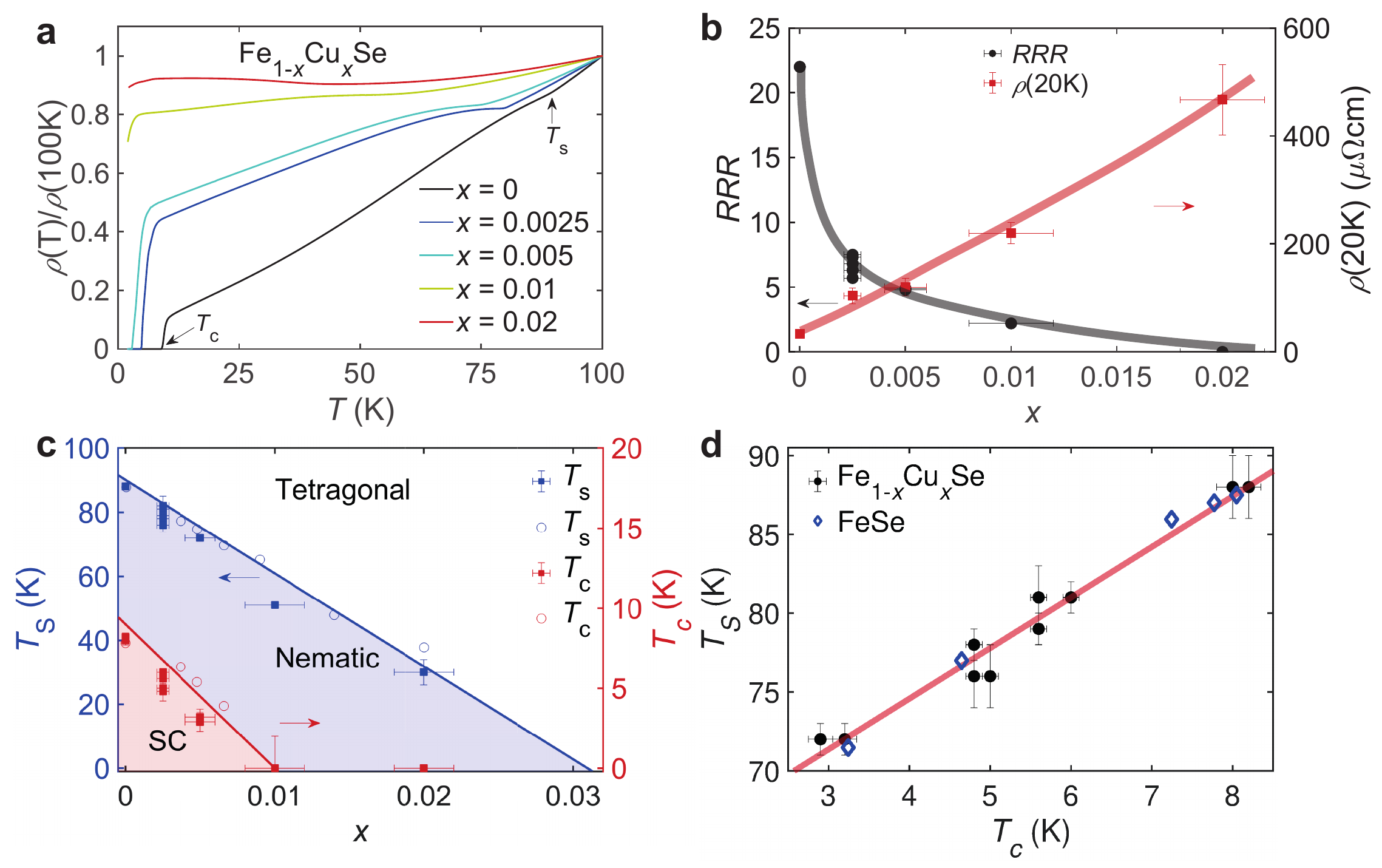}
	\caption{{\bf Transport behaviour for single crystals of Fe$_{1-x}$Cu$_{x}$Se.}
		(a) Temperature dependence of the resistivity as a function of Cu doping showing
an increase in the absolute value of the residual resistivity at low temperatures.
		(b) The evolution of the residual resistivity ratio, $RRR$, (defined as the ratio of the resistivity at 300~K to that at the onset of superconductivity)
 and that of the resistivity at 20\,K with different Cu substitution.
		(c) Temperature-doping phase diagram for Fe$_{1-x}$Cu$_{x}$Se. Open symbols correspond to data previously reported in Ref.~\cite{Gong2021}.
 		(d) The linear dependence between the nematic transition $T_{\rm s}$ and $T_{\rm c}$ caused by the Cu substitution. The blue open diamonds represent data related to FeSe growth using different temperature
 gradients that can lead to the formation of defects in the crystals \cite{Bohmer2016FeSe}.
}
	\label{Fig_x_1}
\end{figure*}

In this paper, we report a detailed study of the electronic behaviour of single crystals of Fe$_{1-x}$Cu$_{x}$Se as a function of different substitution up to $x = 0.02$. The resistivity increases significantly, displaying
bad metallic behaviour at low temperatures, whilst the superconductivity and nematicity are suppressed with increasing Cu substitution.
 From magnetotransport measurements up to 35~T, we detect a reduction of the charge carriers mobilities,
 as compared with FeSe, while the charge carriers densities are hardly affected for low substitution ($x=0.0025$).
 As the hole carriers remain more mobile, the Hall coefficient is positive, in contrast to the negative coefficient
 found in FeSe and in Co-substituted FeSe single crystals.
  The increase in the impurity scattering reduces the magnetoresistance and induces a rather linear resistivity
 at low temperatures inside the nematic phase,
 but the field dependence follows a power law of $\sim H^{1.6}$, similar to FeSe.
 The suppression of superconductivity
 can be described by the Abrikosov-Gor'kov formula for sign-changing $s_{\pm}$ pairing in the presence of non-magnetic impurities \cite{AGformula}.
 The upper critical fields follow similar trends to FeSe, and all curves
 collapsed onto a single line in reduced units of $H_{\rm c2}/T_{\rm c}$, but we cannot
 detect the additional upturn at low temperatures of FeSe when the magnetic field is aligned in plane.
  Under uniaxial compressive strain, the superconductivity is enhanced showing similar trends to those found in bulk FeSe \cite{Ghini2021}.

\section{Experimental details}
Single crystals of different concentrations of
Fe$_{1-x}$Cu$_x$Se, with the nominal
composition varying from $x$ = 0.0025 to 0.02 were grown using the KCl/AlCl$_3$ chemical vapour transport method,
using the same growth conditions for all compositions \cite{Chareev2013,Bohmer2016FeSe}.
For the lowest composition of $x$ = 0.0025 the EDX measurements
indicate a slight variation in composition close to $x$ = 0.0029
and that the ratio between (Fe+Cu)/Se is around 0.95
suggesting an excess of selenium ions or deficit of Fe ions.
Transport studies were performed on more than 20 samples
and their residual resistivity ratio are below 7.5-8, much lower than the values of 25-30 in FeSe.
We observe a larger variation in the resistivity with increasing
$x$ within the same batch due to the inhomogeneous distribution of Cu (see Fig.~\ref{Fig_SM_rhoXXvB}).
In-plane transport measurements ($I ||$($ab$)) at constant temperatures were performed
in a variable temperature cryostat in dc fields up to 35~T at
HFML, Nijmegen for three different crystals (S2, S3, S4) with
the magnetic field applied mainly along the $c$-axis (transverse
magnetoresistance) but also in the ($ab$) conducting plane (longitudinal
magnetoresistance). Low-field
measurements were performed in a 16~T Quantum Design
PPMS. The resistivity $\rho_{xx}$ and Hall $\rho_{xy}$ components
were measured using a low-frequency five-probe technique
and were separated by (anti)symmetrizing data measured in
positive and negative magnetic fields.
Good electrical contacts
were achieved by In soldering along the long edge of the
single crystals, and electrical currents up to 1~mA were used
to avoid heating.
Errors in estimating the exact contact positions and their size result in errors
in the absolute values of resistivity being up to 13\% of the total value.
Strain measurements were performed using a Razorbill cell with the sample glued
to a titanium platform which is compressed. To account for the absolute
value of the strain, correction were made to account for the glue effects,
similar to previous reports on FeSe \cite{Ghini2021}.
Torque measurements were performed using Seiko cantilevers
on a single crystal which was
first tested using X-ray diffraction. Measurements were performed in
constant magnetic field and different temperatures
by rotating the sample between $H||c$ and $H||$($ab$)
to extract the value of the susceptibility anisotropy.

\section{Results and discussion}

\textbf{The transport properties of Cu-substituted FeSe}
Fig.~\ref{Fig_x_1}(a) shows the temperature dependence of resistivity for
single crystals of Cu substituted FeSe for different compositions up to $x = 0.02$.
The resistivity displays a strong anomaly at $T_{\rm s}$, which is the temperature below which
the system enters the electronic nematic phase.
 The stabilization of this electronic phase
can be driven by orbital-ordering effects and electronic correlations,
but the lattice suffers in-plane distortion from a tetragonal to orthorhombic structure  \cite{Coldea2017}.
With Cu doping, the temperature dependence of resistivity changes significantly from its metallic-like behaviour towards an almost invariant temperature dependence below 100~K
caused by the increase in the residual resistivity at low temperature, as expected from Matthiessen's rule.
For compositions higher than $x = 0.02$, the resistivity shows hardly any temperature dependence and the system becomes insulating-like ($d\rho/dT <0$) for $x=0.2$ leading to a factor of 100 increase in
resistivity at low temperature \cite{Williams2009,Li2020}.
The resistivity displays an unusual linear temperature dependence for low Cu substitution of $x=0.0025$
in the normal state which is often a signature of
 a system close to an antiferromagnetic critical point
in the presence of disorder
 \cite{Rosch1999}.

\begin{figure*}[htbp]
	\centering
	\includegraphics[trim={0cm 0cm 0cm 0cm}, width=0.9\linewidth,clip=true]{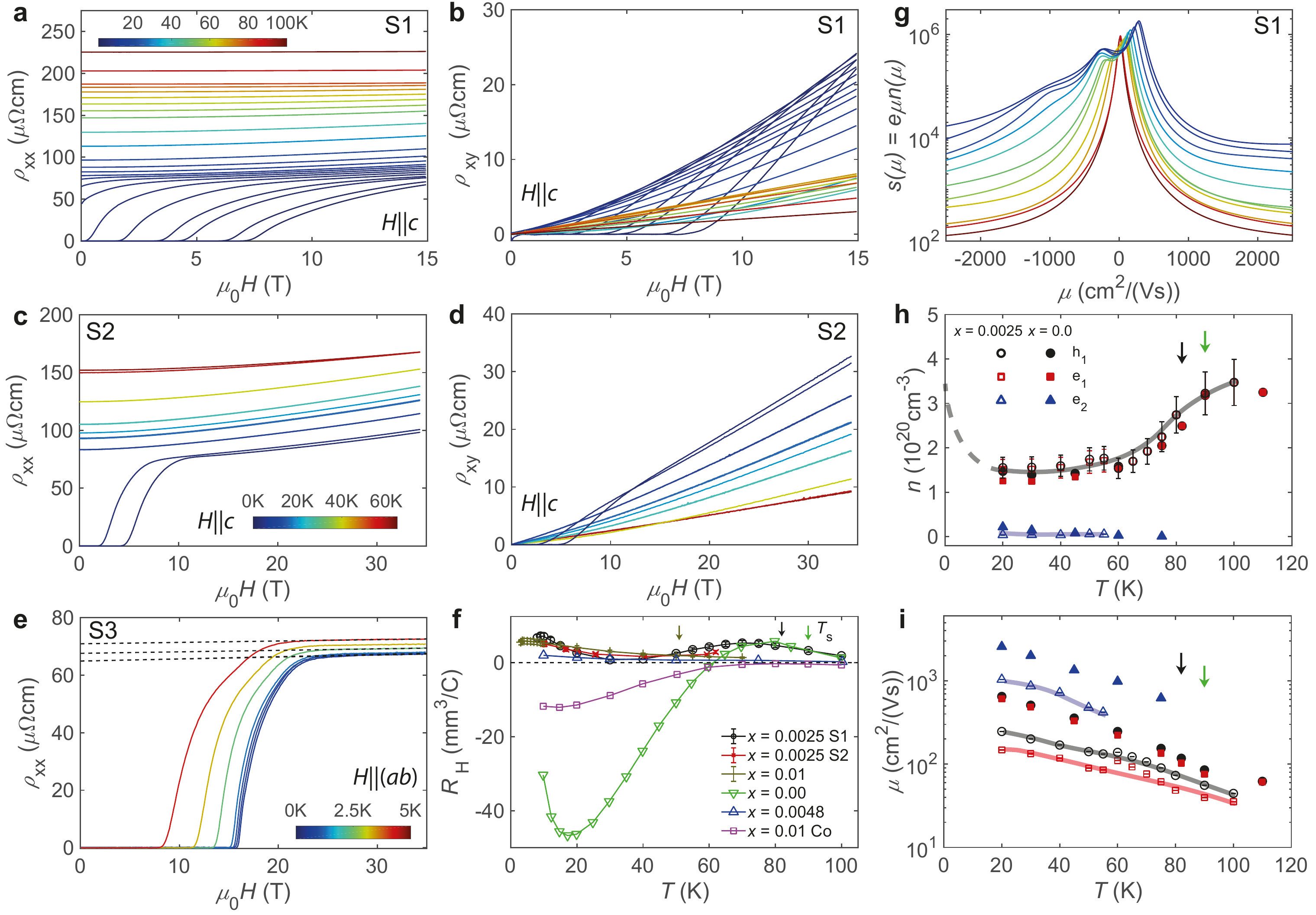}
	\caption{{\bf Magnetotransport results for Fe$_{1-x}$Cu$_{x}$Se with $x=0.0025$.}
		The field dependence of the longitudinal resistivity, $\rho_{xx}$, and Hall resistivity, $\rho_{xy}$, for sample S1 in magnetic fields up to 15~T in (a) and (b),
and for sample S2 measured in magnetic fields up to 35~T in (c) and (d) respectively.
The measurements were performed with the magnetic field along $c$ axis ($H||c$) and at different constant temperatures.
(e) The longitudinal resistivity, $\rho_{xx}$, for the sample S3 was measured with $H||$($ab$) up to 35~T at temperatures below 5~K.
		(f) The temperature dependence of the Hall coefficient, $R_{H}$, for Cu-substituted FeSe
compared with FeSe \cite{Culo2021} (down open triangle),
 $x$ = 0.0048 (up open triangle) from \cite{Gong2021}, and
 Co substitution
 $x$ = 0.001 (open square) from \cite{Urata2016}.
		(g) The mobility spectrum generated from the field sweeps in (a) and (b) using an approach developed in Ref.~\cite{Humphries2016}.
		(h) Carrier densities and (i) mobilities of charge carriers as a function of temperature extracted considering a
compensated two-band model at higher temperatures and a compensated three-band model at lower temperatures below 60~K.
Data for FeSe are taken from \cite{Watson2015b}.
The solid lines is a guide to the eye the expected trend for the apparent carrier number.
The dashed solid line indicates the expected calculated value
 from quantum oscillations data \cite{Coldea2021,Watson2015b}.
This apparent drop in the carrier number inside the nematic phase  is a consequence of the anisotropic scattering, as found for FeSe \cite{Watson2015b}. 
The positions of $T_{\rm s}$ for different compositions are indicated by arrows  of different colours.
		}
	\label{Fig_x_2}
\end{figure*}

\begin{figure*}[htbp]
	\centering
	\includegraphics[trim={0cm 0cm 0cm 0cm}, width=0.8\linewidth,clip=true]{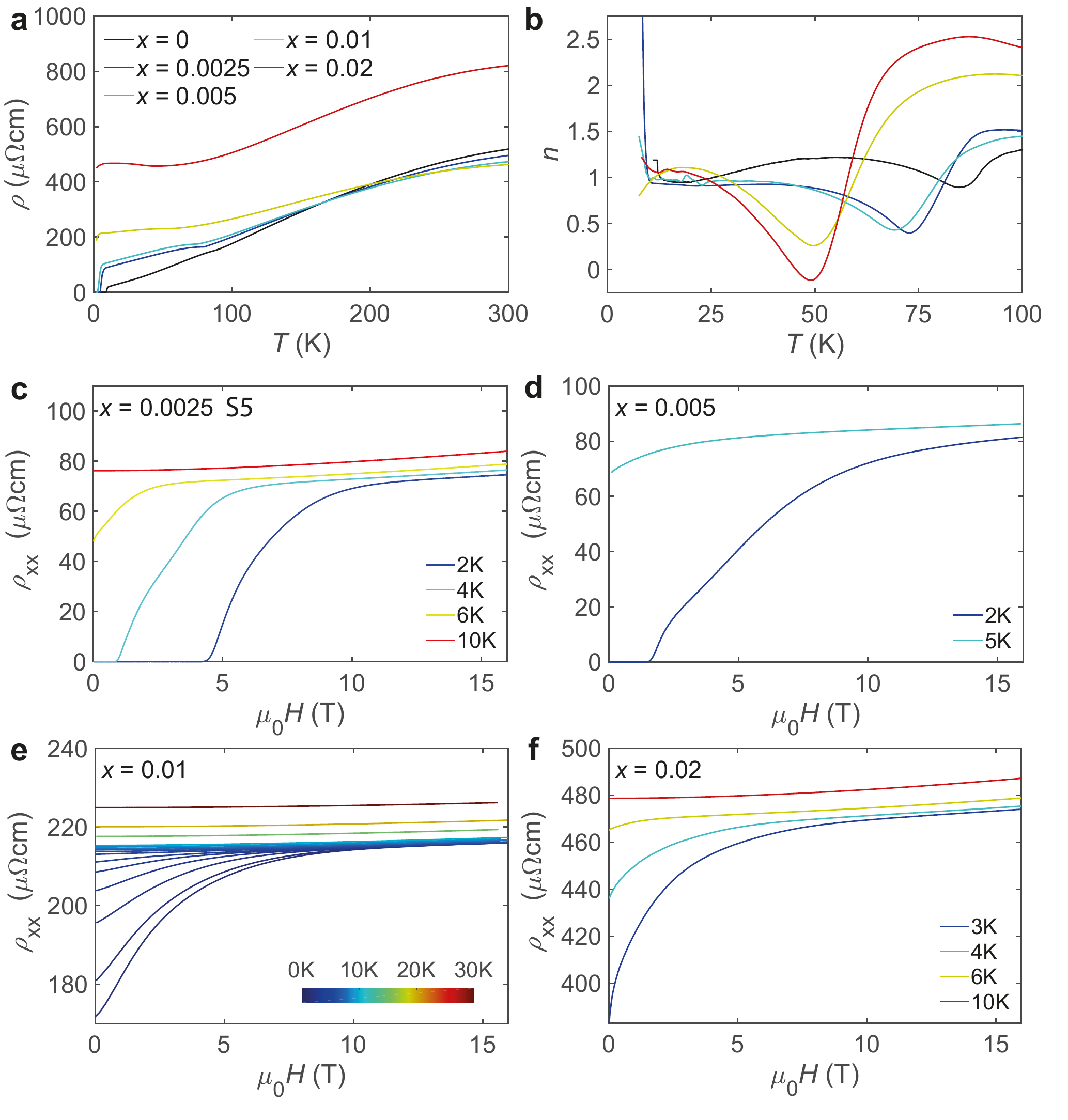}
	\caption{{\bf Transport and magnetotransport of Cu-substituted FeSe.}
		(a) Resistivity temperature dependence for different single crystals with different Cu substitutions, $x$.
		(b) The temperature dependence of the local resistivity exponent, $n$, of the zero field resistivity data in zero-magnetic field, $\rho{(T)} = \rho{(0)} + \alpha T^{n}$.		
		Longitudinal resistivity, $\rho_{\rm{xx}}$, as a function of magnetic field up to 16~T at constant temperatures
for different Cu substitutions: (c) $x = 0.0025$, (d) $x = 0.005$, (e) $x = 0.01$, (f) $x = 0.02$.}
	\label{Fig_SM_rhoXXvB}
\end{figure*}

Together with these striking changes in the temperature dependence of resistivity,
the superconductivity is strongly suppressed and no superconducting state is stabilized down to 0.3\,K  for $x = 0.01$.
The normal-to-superconducting transition becomes very broad,
 $\Delta T \sim 1.5-2$~K, suggesting that the Cu substitution leads to an inhomogeneous electronic distribution, even for very small substitutions.
 The superconducting fraction is reduced significantly with increasing Cu substitution (see Fig.~\ref{Fig_x_1}a) and
 the resistivity just above the onset of superconductivity at 20~K, $\rho_0$(20~K), increases significantly, as shown in Fig.~\ref{Fig_x_1}b.
For higher $x$, the resistivity at low temperatures is reduced in response to the applied magnetic field,
as if this would be a partially superconducting system (see Fig.~\ref{Fig_SM_rhoXXvB}e and f).
This suggests that Cu substitution promotes the formation of isolated clusters of superconductivity that
cannot act as a percolating path in a parallel network resistor to contribute significantly to the total conductivity
and to give rise to zero resistivity at low temperatures. At higher doping, these effects would also lead to more
variation in crystal quality and transport values between individual samples from the same batch
(see also Fig.~\ref{Fig_SM_rhoXXvB}).

The nematic phase is suppressed from 87~K for $x$ = 0 to 29~K for $x = 0.02$, and it becomes
increasingly broader, similar to effects influencing the superconducting transition.
No structural transition or superconductivity is detected in powder samples with $x \geq 3$\% \cite{Huang2010,Li2020}.
As the resistivity temperature dependence weakens with Cu doping, the residual resistivity ratio,
$RRR$, defined here as the ratio of resistivity at 300\,K and onset temperature,
drops significantly from 30 for FeSe \cite{Bristow2020}, as shown in Fig.~\ref{Fig_x_1}b.
This is a measure of the effect of disorder in our system,
similar to previous reports on single crystals \cite{Gong2021,Li2020},
powder samples of Cu-substituted FeSe \cite{Williams2009} and Co and Ni doping \cite{Mizuguchi2009}.
Additionally, the full temperature dependence, from 300~K, for each substitution is shown in Fig.~\ref{Fig_SM_rhoXXvB}a,
with a drastic increase in the high temperature resistivity with larger $x$.
The residual resistivity increases faster with Cu-doping, as compared to Co-doping,
for the same amount of substitution \cite{Urata2016,Mizuguchi2009},
suggesting either that Cu produces a larger impurity scattering potential and/or
may not provide additional charge carriers.
However, all substitutions on the Fe conducting plane lead to the suppression of superconductivity,
in contrast to other iron-based superconductors in which the electron doping with Co normally leads
 to an enhancement in superconductivity \cite{Canfield2010}.
Moreover, in-plane substitution of FeSe has a drastic effect
as compared with isoelectronic substitution outside the plane with sulphur and tellurium for selenium \cite{Coldea2021,Mukasa2021}.

Next, we construct the phase diagram of Fe$_{1-x}$Cu$_{x}$Se and compare it to previous work
on single crystals \cite{Gong2021} and powder samples \cite{Williams2009}, as shown in Fig.~\ref{Fig_x_1}c.
We find a linear correlation between $T_{\rm s}$ and $T_{\rm c}$, similar to studies on single crystals of FeSe grown in different conditions that have a different degree of disorder
\cite{Bohmer2016FeSe}, in Fig.~\ref{Fig_x_1}d.
This trend is likely induced by strong impurity scattering due to reduced values of $RRR$,
in contrast to the behaviour of FeSe tuned by uniaxial strain, where the nematic state is suppressed but the superconductivity is enhanced under compressive strain \cite{Ghini2021}.
Similar trends have been observed also for FeSe thin flakes tuned against the inverse of their thickness, suggesting that disorder
may also play a role in those systems \cite{Farrar2020}.

{\bf Magnetotransport behaviour of Cu-doped FeSe}
The magnetotransport behaviour is studied extensively for the lowest Cu concentration of $x = 0.0025$ to understand the impact of the impurity scattering on the nematic electronic phase in the presence of
anisotropic spin fluctuations \cite{Bristow2020}.
Figs.~\ref{Fig_x_2}a and b show the magnetic field dependence of $\rho_{xx}$ and $\rho_{xy}$ respectively for a sample S1 measured in magnetic fields up to 15~T, and for sample S2 measured up to 35~T
in Figs.~\ref{Fig_x_2}c and d, all with the magnetic field applied perpendicular to the conducting plane ($H||c$). The superconducting transition to normal state becomes rather broad in magnetic field ($\Delta B\sim$ 10~T as compared with FeSe),
 even for this small substitution, and we define zero resistivity in magnetic field as the $H_{\rm c2}$.
Inside the normal state at higher temperatures, longitudinal resistivity, $\rho_{xx}$, has little field dependence,
and $\rho_{xy}$ has rather linear field dependence, as one would expect for a two-band compensated system.
At lower temperatures below 60~K, $\rho_{xy}$ develops a distinctly non-linear behaviour, as shown in Fig.~\ref{Fig_SM_Mobility}d,
but it does not change its sign, as in the case of FeSe inside the nematic phase \cite{Watson2015b}. The magnetotransport close to $T_{\rm c}$ is shown for a sample of each composition in Fig.~\ref{Fig_SM_rhoXXvB}c-f, especially highlighting the reduced superconducting fraction in $x$ = 0.01 and 0.02.

In the low field regime ($\mu_0 H <1$~T), one can extract the Hall coefficient, $R_{\rm H}$, at each temperature from the field dependence of $\rho_{xy}$ assuming a linear dependence of the curves in
Fig~\ref{Fig_x_2}b.
Fig.~\ref{Fig_x_2}f shows the temperature dependence of $R_{H}$ for
Cu and Co substituted FeSe for different values of $x$ \cite{Gong2021,Li2020,Urata2016,Culo2021}. For the lowest Cu doping of
$x=0.0025$, $R_{H}$ remains positive at lowest temperatures, but its magnitude shows weak temperature dependence with a local peak around 75\,K followed by a local minimum around 40\,K. At a higher
substitution of $x = 0.01$, $R_{H}$ remains positive at lowest temperatures increasing slowly by reducing the temperature.
In FeSe, the Hall coefficient changes sign below 70~K with a positive local maximum around 80~K \cite{Watson2015b,Culo2021,Li2020,Bristow2020,Urata2016}.
Different compositions of Cu-doped FeSe systems all display a positive Hall coefficient inside the nematic phase \cite{Gong2021,Li2020} below $T_{\rm s}$, in contrast to Co-doped FeSe where $R_{\rm H}$ remains
negative for all measured samples \cite{Urata2016}.
This suggests that Cu and Co substitution, besides causing strong impurity scattering in the Fe plane, behave differently with respect to the doping of electrons,
with Co being a more significant donor of electrons due to having a negative Hall coefficient \cite{Urata2016}.
If Cu doping would donate extra electrons to the system, one may expect a similar response to the Co doping. On the other hand, in the presence of isoelectronic substitution in FeSe$_{1-x}$S$_{x}$, the Hall
coefficient at low temperatures changes sign and becomes positive for higher $x$, as the distortion of the Fermi surface is suppressed and the anisotropy scattering reduced
\cite{Coldea2021,Culo2021,Li2020,Bristow2020}.
Thus, a combination of the increased dominance of the impurity scattering in relation to the anisotropic scattering in a multi-band system like Cu-doped FeSe can disturb the response of the system and easily affect the Hall coefficient $R_{\rm H}$.

In order to quantify the changes in the magnetotransport in Cu-substituted FeSe
we can simultaneously fit the two resistivity components $\rho_{xx}$ and $\rho_{xy}$
to extract the carrier density and mobilities of the charge carriers for the samples S1 and S2.
Fig.~\ref{Fig_x_2}g shows changes in the mobility spectrum of the charge carriers inside the nematic phase, using a method developed previously in Ref.~\cite{Humphries2016}.
Upon cooling, two peaks develop in the mobility spectrum which are fairly symmetrical around zero, with electrons having a negative mobility and holes corresponding to positive mobilities. As the temperature
decreases inside the nematic phase, both the hole and electrons become more mobile. At lower temperatures inside the normal state, below 60~K,
an additional local peak develops in the mobility spectrum which would correspond to an electron-like carrier.

\begin{figure*}[htbp]
	\centering
	\includegraphics[trim={0cm 0cm 0cm 0cm}, width=0.9\linewidth,clip=true]{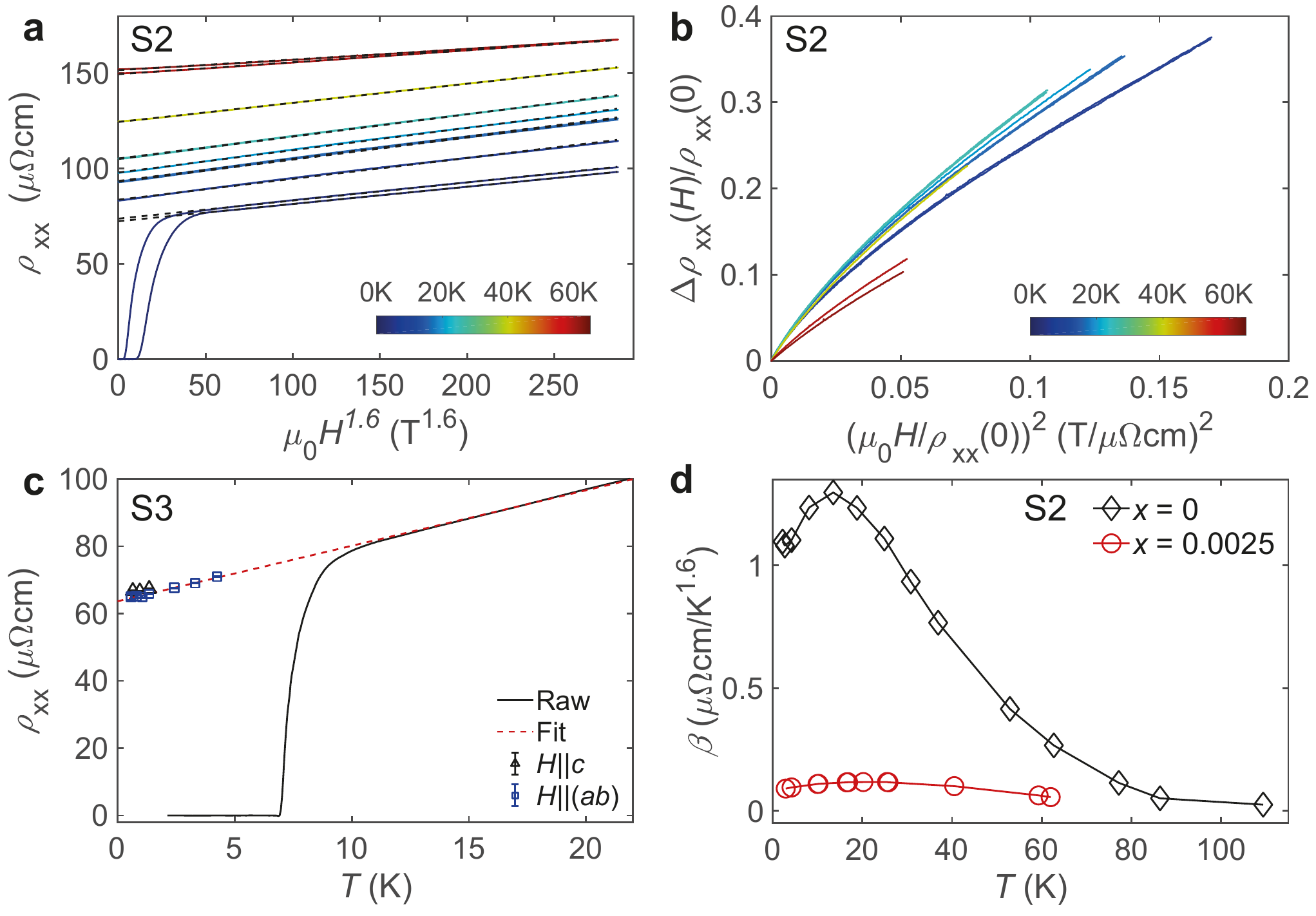}
	\caption{{\bf Normal electronic state of Cu-substituted FeSe with $x=0.0025$.}
	(a) Longitudinal resistivity versus $H^{1.6}$ for sample S2 inside the nematic phase at constant temperatures
(represented in the colour bar) for $H||c$.
The dashed lines are linear fits to these data to find the extrapolated zero field data plotted in panel (c) using $\rho(H,T) = \rho(0,T) + \beta (\mu_{\rm 0} H)^{1.6}$.
	(b) Kohler's scaling applied to the same data as in (a), where $\Delta\rho_{xx} = \rho_{xx}(H,T) - \rho{(0,T)}$.
	(c) The low-temperature zero field resistivity, shown as a solid line, is compared
with the extrapolated values from high field measurements with $H||$($ab$) (blue squares)
and from $H||c$, as shown in (a) (black triangles).
The red dashed line is a fit to the low temperature $H||$($ab$) points.
	(d) The slope of the $H^{1.6}$ magnetoresistance, $\beta$, extracted from (a), as a function of temperature.
The black points are taken from a similar field dependence of FeSe from Ref.~\onlinecite{Bristow2020}.
}
	\label{Fig_normal_state}
\end{figure*}

By using the mobilities from the mobility spectrum as starting parameters,
we performed simultaneous fitting of the two components of resistivity, assuming charge compensation, since low Cu-substitution does not seem
to dope the system with electrons or holes, like Co-doping. Examples of these fits are included in Figs.~\ref{Fig_SM_Mobility}e and f.
Cu was suggested to be an electron dopant \cite{Chadov2010,Huh2021},
but for our low concentration of $x=0.0025$, any excess of electrons or holes is negligible.
Moreover, the Hall coefficient becomes positive by Cu substitution, which is in contrast to Co substitution \cite{Urata2016}.
Therefore, to analyse the magnetotransport data
we employ a compensated two-band model at high temperatures and a three-carrier model at low temperatures,
similar to the approach used in FeSe \cite{Watson2015b,Terashima2016}.
The extracted temperature dependence of both the carrier density and mobilities of Cu-substituted
FeSe with low $x=0.0025$ and its comparison with FeSe are shown in Figs.~\ref{Fig_x_2}h and i, respectively.
The carrier density of the dominant  hole and electron
charge carriers ($n_h$ and $n_{e1}$ )
is reduced inside the nematic state and its behaviour is comparable to that of bulk FeSe \cite{Terashima2016,Watson2015b}.
The apparent reduction in the carrier density, rather than being related to a Fermi surface reconstruction inside the nematic phase,
is likely to reflect the anomalous transport inside the nematic phase caused by anisotropic scattering which still persists in the presence
of Cu substitution and strong impurity scattering \cite{Watson2015b}.
We find that the mobilities of all charge carriers are significantly reduced, as compared with FeSe, by roughly a factor 3, due to the increase in the scattering rate $\mu=e\tau/m^*$, as Cu ions act as strong
scattering centers \cite{Chadov2010}.
An unusual disparity occurs between the hole and electron behaviours at low temperatures, with electron mobilities becoming rather less temperature dependent below 50~K, similar to thin flake devices
\cite{Farrar2022}.
At low temperatures, the impurity scattering dominates the changes in spectrum, as seen in the increased values of $\rho_{0}$, and similar reductions were seen at higher Cu doping at 10\,K \cite{Gong2021}.

{\bf Normal electronic behaviour at low temperatures}
In zero-magnetic field, above $T_{\rm c}$, the resistivity of Cu-doped FeSe displays rather linear resistivity
behaviour with small increase of disorder (Figs.~\ref{Fig_x_1}a and \ref{Fig_SM_rhoXXvB}b), in agreement with previous reports \cite{Gong2021}.
Interestingly, linear resistivity was also found
to be dominant in samples with smaller $RRR$ in FeSe$_{1-x}$S$_x$
and inside the nematic phase \cite{Bristow2020,Zhou2021}.
In order to probe the normal state properties at the lowest temperatures below $T_{\rm c}$,
one can use strong magnetic fields to suppress superconductivity and investigate the extrapolated resistivity in zero magnetic field.
For a precise determination of the normal resistivity below $T_{\rm c}$,
we use two approaches. The first approach is to use the linearly extrapolated
zero-field values of resistivity from the in-plane longitudinal resistivity measurements ($H||I||$(ab)), which eliminates the orbital effects,
as shown for sample S3 in Fig.~\ref{Fig_x_2}e.
The second approach is to identify a suitable field dependence to describe the orbital magnetoresistance for $H||c$;
here, we use a power law of $\sim H^{1.6}$, similar to that found for FeSe \cite{Bristow2020},
and shown in Fig.~\ref{Fig_SM_Mobility}a.
By combining all the extracted data, we find that the normal resistivity data has a linear dependence that extends to the lowest measured temperatures
and gives accurate access to the zero-temperature resistivity, $\rho_{0}$.
Using this parameter, $\rho_{0}$, one can assess the temperature dependence of the local value of the resistivity exponent, $n$,
from the relationship $\rho = \rho_{0} + A T^{n}$ (Fig.~\ref{Fig_SM_rhoXXvB}b).
This confirms the earlier findings
that at the lowest temperatures, inside the nematic phase, $n$ is very close to 1 for Cu-substituted FeSe,
as compared with very clean materials, like FeSe$_{1-x}$S$_x$,
 where a crossover regime to a Fermi liquid behaviour was detected \cite{Bristow2020,Licciardello2019,Reiss2020}.
One important finding is that, in the presence of strong impurity scattering, the linear resistivity region extends
to the lowest temperatures, clearly shown in Fig.~\ref{Fig_normal_state}c, which would be consistent with the existence of strong antiferromagnetic critical fluctuations in the presence of disorder \cite{Rosch2001}.
Similar findings were detected also in thin flakes in the presence of strong two-dimensional fluctuations and disorder \cite{Farrar2022}.

Using the zero-temperature residual resistivity, $\rho_{0}$, one can estimate the mean free path
at low temperatures of the multi-band system FeSe, assuming that its Fermi surface is  formed of
 two-dimensional cylinders with compensated hole and two electron pockets ($k_{\rm F} \sim 0.1 $\AA$^{-1}$ ), such that
$\ell=\frac{\pi c \hbar}{N e^2 k_{\rm F} \rho_{0}}$ \cite{Kasahara2020}.
The mean free path is $\ell\,\sim\,712$~\AA\ for a value of $\rho_{0}\,\sim\,5\,\mu\Omega$cm of FeSe \cite{Bristow2020}
%to 8500~\AA\ for $\rho_{0}\,\sim\,0.4\,\mu\Omega cm$ \cite{Kasahara2020}.
 which is reduced a factor of 10 towards $\ell \sim 55(5)$~\AA~ for the lowest Cu substitution in FeSe
leading to the increase in the residual resistivity.
By using the mobilities extracted from magnetotransport, the scattering time for different pockets of Cu substituted
FeSe varies between 0.6 -1.2~ps,
%leading to an averaged time of 1~ps,
a factor of 3 shorter than in FeSe \cite{Watson2015b}.
With increasing Cu concentration, the residual resistivity increases significantly
reaching the limit $\ell\,\sim\,a$, at which the average distance a
quasi-particle travels between collisions is equal to the interatomic spacing.
The system reaches the Mott-Ioffe-Regel limit \cite{Hussey2004}
where the coherent quasi-particle motion vanishes
and the electron scattering rate $\tau^{-1}$ becomes
comparable to the Fermi energy $E_{\rm F}/\hbar$ \cite{Werman2016}.
With increasing Cu substitution, this limit is reached
for very low substitution at low temperatures, and consequently
systems will become insulating above $x > 0.02$.

\begin{figure*}[htbp]
	\centering
	\includegraphics[trim={0cm 0cm 0cm 0cm}, width=1\linewidth,clip=true]{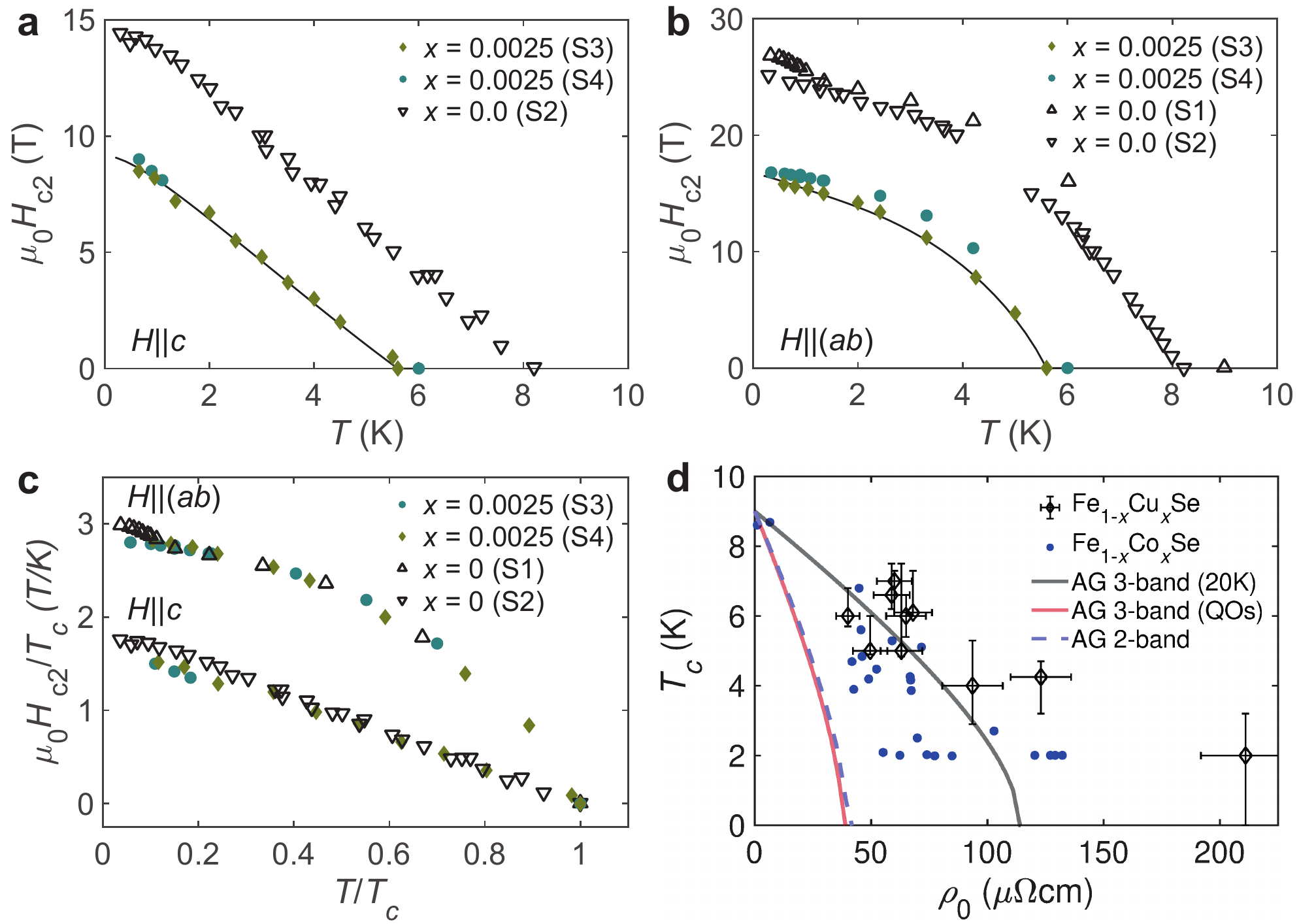}
	\caption{{\bf Superconducting properties of Cu-substituted FeSe. }		
		Upper critical field-temperature phase diagram of Cu-substituted FeSe for $x=$ 0.0025
and FeSe for $H||c$ in (a) and $H||ab$ in (b), respectively.
The position of $H_{\rm c2}$ is defined as the offset of superconductivity at the zero resistance temperature.
Data for FeSe are from Ref.~\onlinecite{MokOk2020}  in (a) and Ref.~\onlinecite{Kasahara2020} in (b).
The data are described by a two-band model using similar parameters to those for FeSe \cite{BristowPhD2020}
but with reduced $\eta$ anisotropy factor.
%the following parameters:
%$T_{\rm c} \sim 5.6$~K, $\lambda_{11}=0.81$, $\lambda_{22}=0$, $\lambda_{12}=\lambda_{21}=0.5$, $\alpha_1=1.6$
%for $H||ab$ and $\alpha_2=0$ $H||ab$.
%$\eta$=0.1 ($v_1$=100.8~meV~\AA, $v_2$=31.9 meV\AA~ for $H||ab$, $v_1=228 meV$\AA, $v_2=72.2 meV$\AA for $H||c$. )
		(c) The reduced upper critical field, $H_{\rm c2}/T_{\rm c}$, versus reduced temperature, $T/T_{\rm c}$,
for two orientations in magnetic field using the data from (a) and (b).
		(d) The suppression of superconducting transition temperature, $T_{\rm c}$, by the impurity scattering
expressed by the zero temperature residual resistivity $\rho_{0}$ of Cu-substituted FeSe (open diamonds)
and Co-substituted FeSe (solid circles) \cite{Urata2016}.
The solid red line is a 3-band AG formalism as seen in \cite{Urata2016} using
parameters from quantum oscillations and previous magnetotransport studies \cite{Watson2015a,Watson2015b}.
The effective masses used are: $m_{h1}=4.5~m_e$, $m_{e1}=7~m_e$, $m_{e2}=1.5~m_e$.
The carrier densities used for the three band model are
estimated from quantum oscillations at the lowest temperatures (low $T$) (red solid line):
$n_{h}=3.75 \times 10^{20}$ cm$^{-3}$
$n_{e1}=4.33 \times 10^{20}$ cm$^{-3}$ and
$n_{e2}=0.78 \times 10^{20}$ cm$^{-3}$
and those from magnetotransport analysis at $T$ = 20~K \cite{Watson2015b} (grey solid line) are:
$n_1=1.45 \times 10^{20}$ cm$^{-3}$,
$n_2=1.25 \times 10^{20}$ cm$^{-3}$,
$n_1=0.22 \times 10^{20}$ cm$^{-3}$.
The dashed line is a 2-band AG formalism reported previously in Ref.~\cite{Urata2016}.
	}
	\label{Fig_Hc2}
\end{figure*}

The classic magnetoresistance and symmetry conditions in a tetragonal system
lead to a quadratic dependence of the electrical resistivity $\Delta \rho/\rho$ on magnetic field $(\mu_{0} H)^2$ in the low-field limit \cite{Hu2008}.
Studies on FeSe$_{1-x}$S$_x$ detected quadratic dependence of magnetoresistance
only outside the nematic phase, whereas inside
the nematic phase the in-plane distortion of the
Fermi surface and the anisotropic spin fluctuations
could lead to a different power law dependence, which is found to be $\sim H^{1.6}$ \cite{Bristow2020}.
We find that a similar power law describes
also the weakly Cu substituted FeSe for $x=0.0025$, as shown in Fig.~\ref{Fig_normal_state}a.
Interestingly, by comparing the amplitude of this power law $H^{1.6}$ for
Cu-substituted FeSe and FeSe,
we find the change in magnetoresistance over the same field regime to 35~T
 is significantly suppressed by a factor of 10, as compared with FeSe,
 which is close to the expected change due to the reduction in mobilities by a factor 3,
as shown in Fig.~\ref{Fig_normal_state}d.

\begin{figure*}[htbp]
	\centering
	\includegraphics[trim={0cm 0cm 0cm 0cm}, width=0.8\linewidth,clip=true]{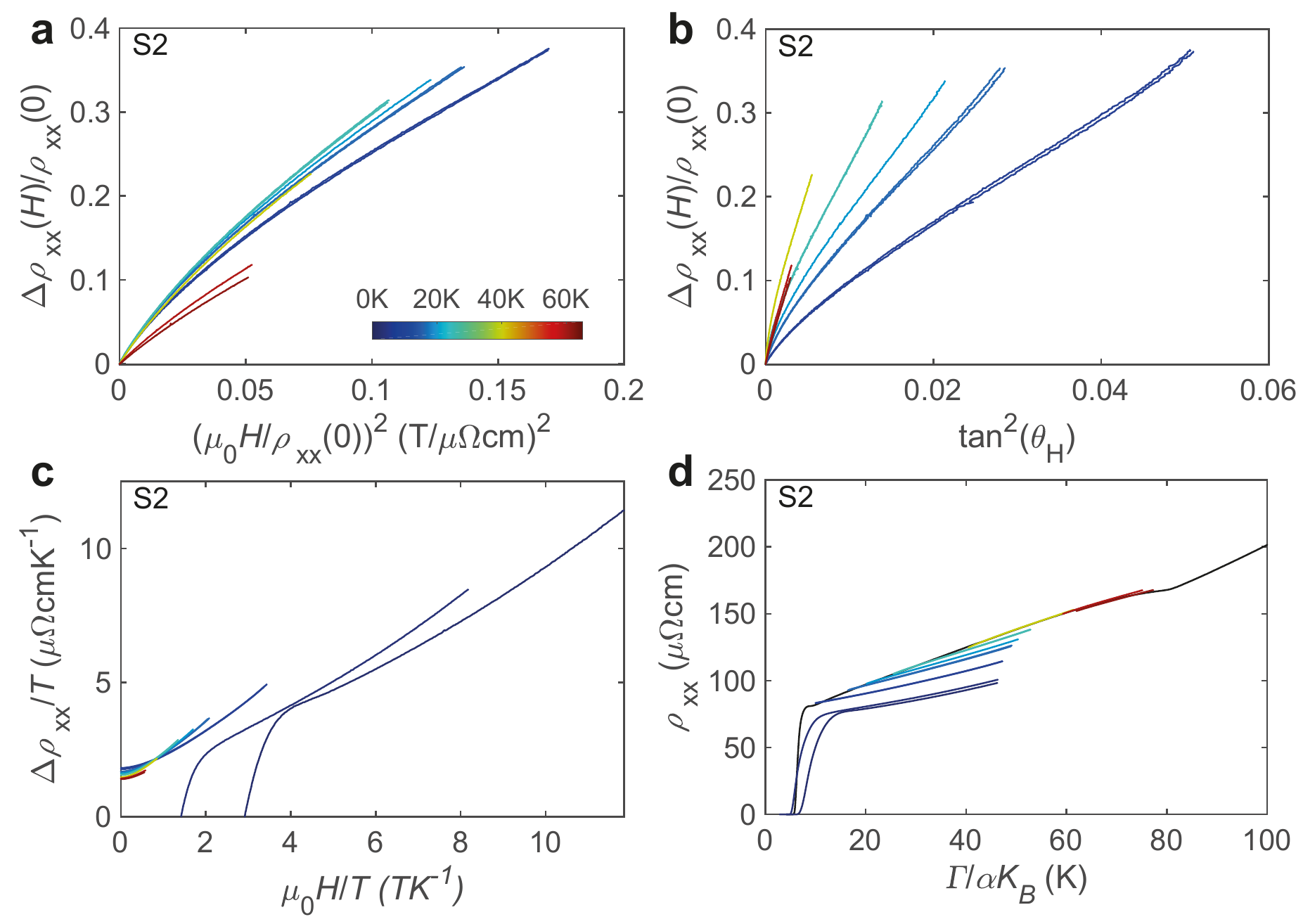}
\caption{{\bf Scaling of the longitudinal magnetoresistance of Cu-substituted FeSe with $x = 0.0025$ ($H||c$).}
	(a) Kohler's rule scaling $\Delta \rho_{xx}(\it{H})/ \rho_{xx}(\rm{0}) \sim (\mu_{\rm{0}}\it{H}/\rho_{\rm{xx}}(\it{H}\rm{0}))^{\rm{2}}$,
where $\Delta\rho_{\rm{xx}}(\rm{H,T})=\rho_{\rm{xx}}(\rm{H,T})-\rho_{\rm{xx}}(\rm{0,T})$.
Kohler's rule violation is evidence that electrical transport is not governed by a single scattering time.
(b) Modified Kohler's rule scaling using the Hall angle, $\tan \theta_{\rm H}=\rho_{xy}/\rho_{xx}$.
(c) $H -T$ scaling of $\Delta\rho_{\rm{xx}}(H)/\rho_{\rm{0}} \sim \mu_{\rm{0}}\it{H}/\it{T}$,
where $\Delta\rho_{\rm{xx}}(H)=\rho_{\rm{xx}}(H)-\rho_{\rm{0}}$ and $\rho_{\rm{0}}$
is the zero-temperature zero-field resistivity.
(d) Energy scaling of resistivity as $\Gamma$, where $\Gamma=\alpha{}\it{k}_{\rm{B}}\it{T}\sqrt{1 +
(\beta/\alpha)^{\rm{2}}(\mu_{\rm{B}}\mu_{\rm{0}}\it{H}/(\it{k}_{\rm{B}}\it{T}))^{\rm{2}}}$ using $\alpha=1$ and $\beta=1$.
 Our data do not follow the proposed energy scaling, similar to FeSe$_{1-x}$S$_{x}$ \cite{Bristow2020}.
This magnetoresistance scaling was used to describe the antiferromagnetic critical region in BaFe$_2$(As$_{1-x}$P$_x$)$_2$ \cite{Hayes2016}.
 }
	\label{Fig_SM_ZP2_ScalingLaws}
\end{figure*}

The Cu substitution would be expected to
impose a single dominant impurity scattering process, such that the Kohler’s rule of magnetoresistance
would be obeyed, $\Delta \rho_{xx}$/$\rho_{xx}(0) \sim (\mu_0 H/\rho_{xx}(0))^2$.
We observe deviations from Kohler’s rule despite the fact
that the impurity scattering increases significantly, as shown in Fig.~\ref{Fig_normal_state}b.
As the resistivity shows rather linear temperature dependence,
we have tested other magnetoresistance proposals, including a modified Kohler's rule
 (Fig.~\ref{Fig_SM_ZP2_ScalingLaws}b), $H$-$T$ scaling (Fig.~\ref{Fig_SM_ZP2_ScalingLaws}c),
 and an energy scaling (Fig.~\ref{Fig_SM_ZP2_ScalingLaws}d) which successfully described the
 antiferromagnetic critical region in BaFe$_2$(As$_{1-x}$P$_x$)$_2$ \cite{Hayes2016}.
Interestingly, this energy scaling has been
suggested to also be fulfilled for dirty FeSe$_{0.82}$S$_{0.18}$ with rather linear resistivity
where the magnetoresistance curves do not cross each other \cite{Licciardello2019MR},
 as opposed to the clean limit in which quantum oscillations were observed \cite{Coldea2019}.
The various proposals for magnetoresistance scaling cannot describe the behaviour of
Cu substituted FeSe (see Fig.~\ref{Fig_SM_ZP2_ScalingLaws}), similar to FeSe$_{1-x}$S$_x$
inside the nematic phase \cite{Bristow2020}.

{\bf Upper critical fields}
Figs.~\ref{Fig_Hc2}a and b show
the temperature dependence of the upper critical magnetic field, $H_{\rm c2}$,
for two single crystals S3 and S4 with $x = 0.0025$ for two different orientations in magnetic field.
These results are compared to those of bulk FeSe \cite{MokOk2020,Kasahara2020}.
At the lowest temperatures, the upper critical field of FeSe shows
an unusual upturn in critical field for $H||$($ab$) which was associated to the stabilisation of an FFLO state \cite{Kasahara2020},
magnetic field-induced transitions \cite{MokOk2020} or it is just the manifestation of multi-band effects on
the upper critical field \cite{BristowPhD2020}.
With the Cu-substitution, we introduce a significant amount of disorder and we do not identify
any further upturn in $H_{\rm c2}$ for $H||$($ab$), similar to the case of thin flakes of FeSe \cite{Farrar2020}.
Due to the slight variation in the values of $T_{\rm c}$
there are shifts between the $H_{\rm c2}$ dependencies for different samples.
By plotting the upper critical fields curves in reduced units $H_{\rm c2}/T_{\rm c} $
versus temperature reduced units, $T/T_{\rm c}$ (using the offset temperature of superconductivity),
all curves collapsed onto a single dependence, as shown in Fig.~\ref{Fig_Hc2}c.
This indicates that the superconductivity pairing mechanism does not change
significantly for low Cu substitution and the low temperature upturn for $H||$($ab$) is smeared out.
The upper critical field behaviour of Fe$_{1-x}$Cu$_{x}$Se could be a consequence of
changes in pairing due to presence of impurities that can promote intraband pairing over interband pairing.
Interestingly, in the case of single crystals of FeSe with small amount of disorder ($RRR$ reduced only to 12
and $T_{\rm c}$ reduced from 9.1 to 7.2~K)
the slight upturn in upper critical field $H||$($ab$)
was found to be robust \cite{Zhou2021}. This observation led to suggestions
that the high-field phase of FeSe is not a conventional FFLO state \cite{Zhou2021}.
However, for enhanced disorder, as in our study,
 these features are smeared out, but the temperature dependence can still be
described by a similar multi-band model suitable for FeSe, but with a reduced velocity anisotropy \cite{BristowPhD2020}.

In order to quantify the behaviour of the upper critical field,
we first use the standard three-dimensional Werthamer-Helfand-Hohenberg (WHH) model to estimate the low temperature orbitally limited critical fields
and assess the value of the Maki parameter,
similar to previous studies on FeSe thin flakes \cite{Farrar2020,BristowPhD2020}.
 Orbital pair breaking alone accounts
 for the temperature dependence of $H_{\rm c2}$ for $H||c$,
 with a slope of $H'_{\rm c2} \sim $ -1.7(1)T/K, similar to FeSe \cite{BristowPhD2020}.
 However, when the magnetic field is aligned along the conducting ($ab$) plane,
 a Pauli pair breaking contribution has to be included which reduces the orbital-limited critical field.
For $H||(ab)$, the slope in the low field regime is $H'_{\rm c2} \sim $ -4.7(1)T/K
which gives an orbital value of 19.2~T, which is larger than the experimental value
15.8~T at 0.6~K, due to the Pauli paramagnetic effects.
For FeSe, the expected Pauli paramagnetic limit field, based on the values of different band gaps, can vary between 4.8~T to 28~T \cite{BristowPhD2020}, for which
the Maki parameter would vary between 1-1.5 for the hole and electron pocket,
increasing to 5.7 for smallest gap \cite{BristowPhD2020}.

In order to describe the temperature dependence of the upper critical field
we use a two-band model in the clean limit,
as the coherence length of Cu substituted FeSe with $x=0.0025$
is 3.4~nm and $\xi \ll \ell$.
The model chosen to describe the temperature dependence of $H_{\rm c2}$
are similar to those employed for thin flakes and bulk FeSe \cite{Farrar2020,BristowPhD2020}.
% The extracted Maki parameter $\alpha_M$ is 2.4 for thick flakes, close to the value of $\alpha_M$ = 2.1 found for bulk single crystals.
Figs.~\ref{Fig_Hc2}a and b show the fitted data
described by the following coupling parameters for both orientations using
%$T_{\rm c} \sim 5.6$~K,
$\lambda_{11} = 0.81$, $\lambda_{22}=0$, $\lambda_{12} = \lambda_{21} = 0.5$, $\eta = 0.025$
with $\alpha_1 = 1.6$ and $\alpha_2 = 0$ for $H||ab$.
% ($v_1$=100.8~meV~\AA, $v_2$=31.9 meV\AA~ for $H||ab$, $v_1=228 meV$\AA, $v_2=72.2 meV$\AA for $H||c$. )
We find that the coupling parameters are similar to FeSe,
whereas the velocity anisotropy, $\eta$, is
slightly larger  than for the highly anisotropic FeSe \cite{BristowPhD2020}.
This suggests that the Cu substitution and the increase in impurity
scattering is smearing the superconducting gap.

One can evaluate the changes in $T_{\rm c}$ against the residual resistivity of FeSe substituted with Cu, and compare the results with those reported for Co substitution \cite{Urata2016} and
previous Cu-substituted FeSe \cite{Gong2021}.
Normally, the suppression of $T_{\rm c}$ by nonmagnetic impurities in iron-based superconductors with $s_{\pm}$ sign reversal superconducting states would obey
the Abrikosov-Gor’kov (AG) formula \cite{AGformula} similar to a magnetic impurity in a single-band BCS superconductor.
A model specific to FeSe implies a full suppression of superconductivity when the residual
resistivity is close to 4~$\mu \Omega cm$
\cite{Urata2016}. In order to estimate the suppression of the $T_{\rm c}$ due to impurity scattering we assume a three-band model of the Fermi surface of FeSe,
considering the presence of the additional small
electron pocket from the mobility analysis. The effective masses and charge carrier densities are taken from previous quantum oscillations and magnetotransport studies at low temperatures \cite{Watson2015a,Watson2015b}, as detailed in the caption of Fig.~\ref{Fig_Hc2}.
With these parameters, one can estimate an average scattering time, considering the contributions of different bands to the
 total conductivity in a parallel resistor network.
Fig.~\ref{Fig_Hc2}d shows the variation of the critical temperature with residual resistivity for Cu and Co doping using current data and
previous reported results from Ref.~\onlinecite{Urata2016}.
Indeed, we find that the suppression in superconductivity obeys the AG formula
using parameters from magnetotransport data (the carrier density and mobility from 20~K in Figs.~\ref{Fig_x_2}h and i \cite{Watson2015b}),
while a stronger suppression occurs using the low-temperature parameters from quantum oscillations \cite{Watson2015a}.
The decrease in carrier densities observed in the mobility at 20~K
is likely to be a consequence of very anisotropic scattering inside the nematic phase
and captures the suppression of $T_{\rm c}$ better than the low temperature results, where the isotropic scattering would be more dominant.
This finding is consistent with a $s_{\pm}$ pairing symmetry in FeSe and emphasizes the importance
of the anisotropic scattering  inside the nematic phase and its effect on the anomalous magnetotransport.
 Such a pairing symmetry is in agreement with the finding from scanning tunnelling microscopy of FeSe \cite{Sprau2017}.

\begin{figure}[htbp]
	\centering
	\includegraphics[trim={0cm 0cm 0cm 0cm}, width=1\linewidth,clip=true]{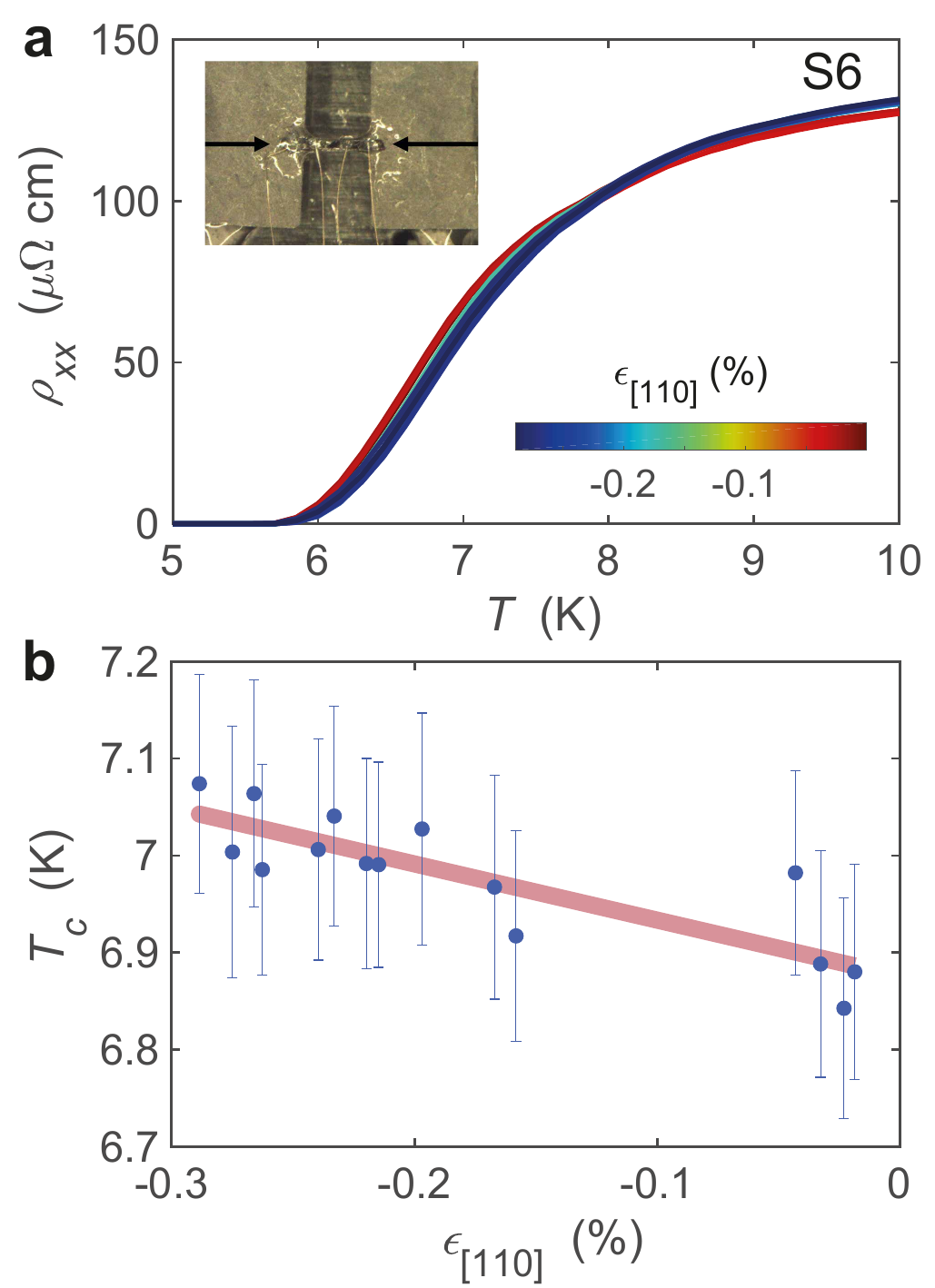}
	\caption{{\bf Applied uniaxial strain of Cu-substituted FeSe with $x=0.0025$.}		
(a) Temperature dependence of the resistivity near the superconducting to normal transition under
compressive uniaxial strain along the [110] direction which is the current direction.
 As the sample (S6) in the insert is glued to a titanium platform,
 it will be exposed to an additional small tensile strain, $\epsilon_{glue} \sim 0.02\%$
 and an additional correction of $\epsilon_{[110]} \sim 0.7 \epsilon_{applied}$
is made to account for the strain transmission to the sample, similar to FeSe \cite{Ghini2021}.
(b) Variation of superconducting transition temperatures, $T_{\rm c}$, defined as the temperature at the midpoint between the
onset and offset (peak in the derivative) under applied compressive strain.
The superconducting state is clearly enhanced under compressive uniaxial strain, similar to FeSe \cite{Ghini2021}.
Solid line is a guide to the eye.}
	\label{Fig_CuFeSe_strain}
\end{figure}

{\bf Effect of applied uniaxial strain}
Fig.~\ref{Fig_CuFeSe_strain}a shows the effect of the compressive strain along the [110] tetragonal direction
on the transport behaviour of a single crystal of Cu-substituted FeSe
close to the superconducting-normal transition.
In the normal state the resistivity increases under compressive strain,
and all resistivity curves, independent from the amount of uniaxial compressive strain applied,
cross around 8.0(1)~K in the vicinity of the superconducting onset temperature transition around 8.2~K.
This behaviour is rather similar to FeSe but the crossing point is lower as the $T_{\rm c}$ is smaller and the transition widths are larger \cite{Ghini2021}.
If this behaviour originates from the possible formation of Griffiths-like phases in the vicinity of the superconducting to insulating transition under strain,
one can expect that the additional Cu-induced disorder
would further enhance these effects \cite{Ghini2021,Reiss2021}.
Additionally, in Fig.~\ref{Fig_CuFeSe_strain}b, the superconducting transition, $T_{\rm c}$,
defined as the peak in the derivative, increases slightly with compressive strain, similar to FeSe \cite{Ghini2021}.
Our findings emphasize that for low Cu substitution
the overall response to strain of FeSe remains unaltered, despite
the significant increase in the impurity scattering which leads to the increased width of the transition,
increase in resistivity and the reduction of the transition temperature.

{\bf Discussion}
Our study has focused on understanding the role of the Cu substitution in FeSe,
 in particular on the changes of the electronic and superconducting behaviour that occur for small substitutions.
Normally, the substitution with a metal-transition ion like Co adds additional electrons into the Fe plane and affects its spin state,
but Cu substitution acts mainly as a source of impurity, at least for low substitutions \cite{Pelliciari2021}.
As a result, the superconductivity is suppressed and resistivity increases
but the field-induced magnetotransport power law and the upper critical field behaviour inside the nematic phase have surprisingly similar trends to FeSe.
If Cu is an electron dopant we should expect a similar behaviour to the Co substitution, which we do not detect for
low amount of substitution.
Recent ARPES studies find  that only large Cu substitution in FeSe
potentially dope the system with electrons \cite{Huh2021}, and
drive it towards a more correlated Mott-like state \cite{Skornyakov2021},
 as long as the surface of the sample is not charging during photoemission studies of non-metallic samples.

Superconductivity of FeSe is strongly suppressed in the presence of Cu and Co \cite{Gong2021,Urata2016},
as compared with the isoelectronic substitution of S and Te \cite{Coldea2021}.
Spin fluctuations, with potential orbitally-dependent character, are likely
responsible for the superconducting pairing in the FeSe family, and remain unchanged with increasing S, but are strongly suppressed by Co substitution \cite{Baek2020}.
This would lead to a $s_{\pm}$ pairing mechanism supporting both nodeless and nodal states and would stabilize
 highly anisotropic superconducting gaps on both electron and hole pockets \cite{Sprau2017}.
The presence of such nodes or gap minima are found in crystals of higher quality only, with a
small amount of disorder being sufficient to smear out the small gap in more isotropic lower quality crystals of FeSe \cite{Sun2018}.
The Cu substitution, which introduces a significant amount of impurity scattering, leads to a suppression of velocity anisotropy,
as detected from the parametrization of the upper critical field (Fig.~\ref{Fig_Hc2}).
The suppression of superconductivity with Cu and Co impurity substitution is consistent
with sign changing $s_{\pm}$ pairing, described by the AG formula.
However, the effect of anisotropic scattering can affect the apparent drop in the carrier density inside the nematic
phase \cite{Watson2015b} and consequently it needs to be taken into consideration in an extended AG formula.

Next, we discuss the effect of impurities induced by Cu in FeSe as compared with those introduced by electron irradiation (2.5 MeV) \cite{Teknowijoyo2016}.
Interestingly, the superconducting transition temperature is slightly enhanced by the point-like disorder induced by irradiation,
and it was suggested that the irradiation-induced Frenkel defects enhanced the pair interaction, in turn enhancing the spin fluctuations \cite{Teknowijoyo2016}.
This increase $T_{\rm c}$ of 0.4~K leads to a decrease in the nematic temperature $T_{\rm s}$  of 0.9~K.
The small change in $T_{\rm c}$ could be related to the relatively low concentration of radiation defects, as well as
the location of these defects in the conducting Fe layers or outside in the van der Waals gaps of FeSe.
Therefore, the small changes in the transition temperatures with electron irradiation
do not follow the trends observed by Cu-doping, but are closer to the behaviour of FeSe substituted with S outside the Fe plane
as well as FeSe under the effect of the small applied pressure and uniaxial strain \cite{Terashima2016,Ghini2021}.
The variation in the response of superconductivity to different kinds of impurities
suggests that the pairing mechanism has a strong three-dimensional dependence
that needs to be taken into account, as suggested by $k_z$ dependent ARPES studies \cite{Kushnirenko2018}.

\begin{figure*}[htbp]
	\centering
	\includegraphics[trim={0cm 0cm 0cm 0cm}, width=0.8\linewidth,clip=true]{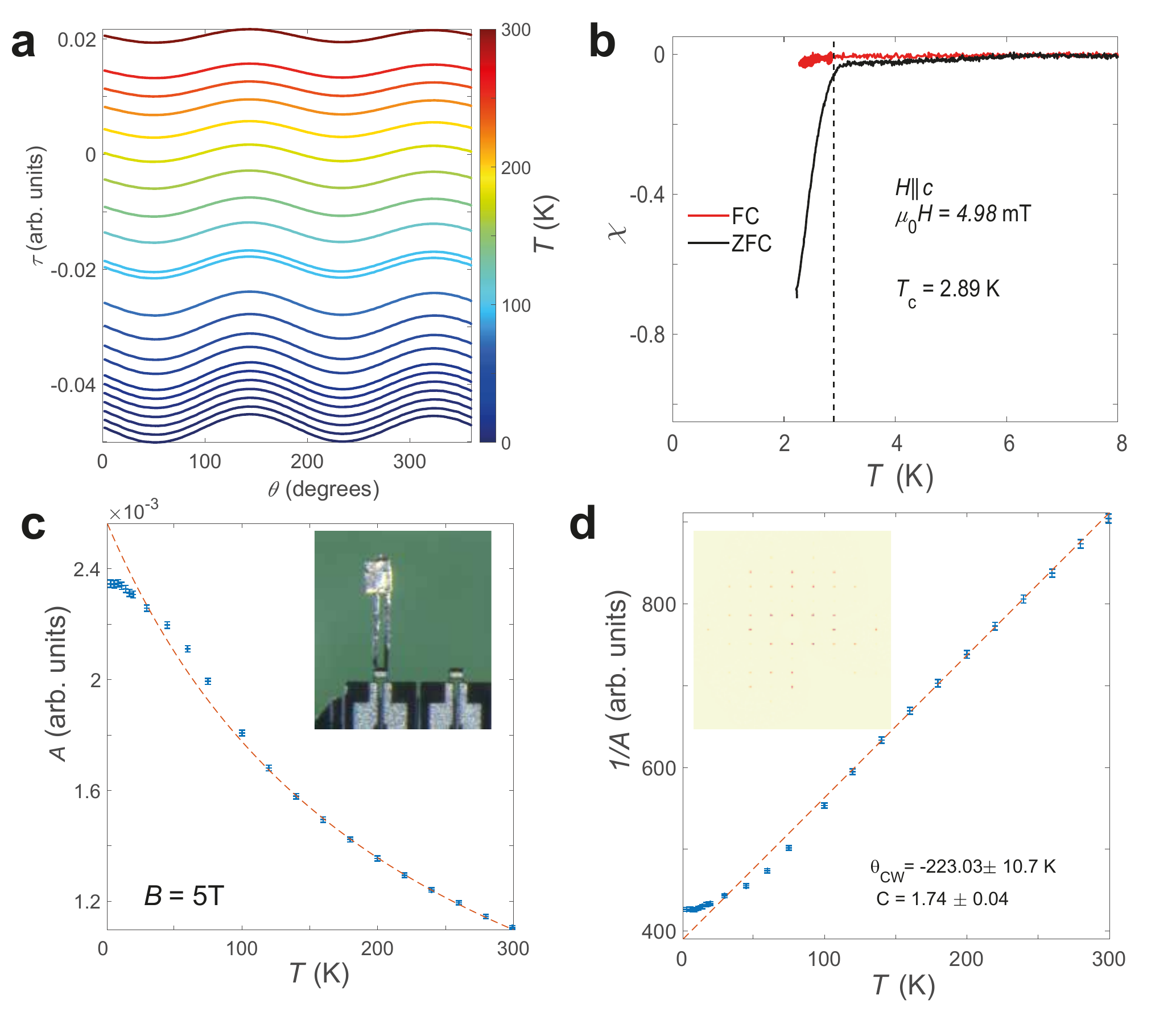}
	\caption{{\bf Magnetic properties of Fe$_{1-x}$Cu$_{x}$Se.}
		(a) Angular dependence of magnetic torque in 5~T at different constant temperatures for a small single crystal with $x$ = 0.02.
 %(SS12)
(b) Temperature dependence of magnetization for a different single crystal with $x$ = 0.01
%(FeSeCu1SS9M1)
 which shows a $T_{\rm c} \sim 2.89$~K but
for only a small fraction of the sample (the diamagnetic factor for this sample is $N$ = 0.94).
(c) The extracted torque amplitude proportional to the spin susceptibility anisotropy versus temperature
for this crystal shown mounted on the piezocantilever (as shown in the inset).
The dashed line is a fit to a Curie-Weiss temperature dependence.
(d) The inverse of torque amplitude versus temperature to extract the Curie-Weiss temperature from the linear fit intercept of $\theta_{CW}$ = -223(10)~K.
The inset shows the X-ray diffraction spectra used to extract the lattice parameters for the
crystal used in torque studies in (a) with a tetragonal
symmetry and lattice parameters
%(SS12 x~2\% XZ1)
$a$ = $b$ = 3.7770(3) \AA , $c$ = 5.5160(6) \AA .
These values are similar to those of FeSe $a$ = $b$ = 3.7651 \AA , $ c$ = 5.5178 \AA ,
and are in agreement with previous reports on powder samples \cite{Huang2010,Williams2009}.
}
	\label{Figure_CuFeSe_torque}
\end{figure*}

As FeSe has been suggested to be close to a magnetic instability,
one could expect that weak repulsive impurities can promote short-range magnetism, with the induced magnetization cloud modified by the orbital selectivity \cite{Martiny2019,Chadov2010}.
To investigate whether Cu substitution in FeSe
can lead to formation of local magnetic moment,
we performed a torque magnetometery study that reveals the behaviour of the susceptibility anisotropy
for $x$ = 0.02, shown in Fig.~\ref{Figure_CuFeSe_torque}.
The response of the system is that of a paramagnetic system following a Curie-Weiss behaviour and
no well-defined anomaly of a long range magnetic order is found, see Figs.~\ref{Figure_CuFeSe_torque}a, c, and d.
However, the superconducting fraction is reduced by the increase in Cu substitution (Fig.~\ref{Figure_CuFeSe_torque}b)
and eventually only random superconducting puddles will be present, as indicated by the finite resistivity which is affected by magnetic fields as one may expect in a superconducting system (Fig.~\ref{Fig_SM_rhoXXvB}e and f).
However, the appearance of an insulating phase by Cu substitution in FeSe displays the hallmark of a strong impurity potential
which can slow the fluctuating magnetic spins and can lead to the enhancement of the ordered magnetic moment.

The precise power law of the temperature dependence of resistivity is often influenced by the nature
of the critical fluctuations, its dimensionality, the presence of disorder and its proximity to quantum critical points.
 At lowest temperatures, the resistivity of FeSe$_{1-x}$S$_x$ displays a Fermi-liquid-like behaviour whereas inside
 the nematic phase a regime of linear resistivity was associated to the presence of spin fluctuations \cite{Bristow2020,Reiss2020}.
 Interestingly, in FeSe samples with larger amounts of disorder (smaller $RRR$ values) the resistivity seem
 to display linear behaviour at the lowest temperatures. This behaviour was also found
 in FeSe crystals with small $RRR$ values \cite{Zhou2021,Bristow2020}, Cu-substituted FeSe and thin flakes of FeSe \cite{Farrar2022}.
In the vicinity of an itinerant antiferromagnetic quantum critical point, as proposed for FeSe \cite{Grinenko2018},
 the resistivity is strongly affected by small amounts of disorder.
Strongly anisotropic scattering due to spin fluctuations  in the presence of disorder would generate
an anomalous temperature dependence of resistivity with exponents varying between $n=1$ and 1.5  \cite{Rosch1999}.
We find in our data that the linear $T$ resistivity persists over a large temperature regime,
to lowest temperatures but it is superimposed on a large background caused
by the resistance of the channel dominated by impurity scattering.
For higher Cu substitution, the Mott-Ioffe-Regel limit, in which the usual description of a metal in terms of ballistically propagating quasiparticles is no longer valid,
and this can lead to suppression of spin fluctuations and enhanced local magnetism
that would lead to the disappearance of superconductivity.

\section{Summary}

In summary, we present a detailed study of the effect of impurity scattering
induced by Cu substitution on the electronic
and superconducting properties of FeSe.
We investigate in detail
for a very low concentration ($x$ = 0.0025)
the suppression of superconductivity, the upper critical fields,
magnetotransport in high magnetic fields
and the response to applied uniaxial strain.
Both the suppression of superconductivity
and the behaviour of the upper critical field
can be accounted for by taking into account the multi-band effects
in the sign-reversal symmetry ($s_{\pm}$) of the order parameter.
Anisotropic scattering affects the apparent charge carrier densities inside the nematic phase,
and in turn this influences the
suppression of $T_{\rm c}$ described by the AG formula.
The magnetotransport studies reveal suppression of the charge mobilities due to the increases in the scattering time
in Cu-substituted FeSe and the reduction in the magnetoresistance. However, the power law of resistivity seems to
have a similar field dependence to FeSe, suggesting the electron-electron collisions remain unaltered by the electron-impurity collisions for low Cu substitutions.
Our study also raises questions about the universality of impurities, in-plane doping,
the effect of irradiation and growth conditions in suppressing superconductivity of FeSe.

In accordance with the EPSRC policy framework on research data, access to the data will be made available
from ORA (DOI to be provided on acceptance).

\section{Acknowledgements}
We thank Andreas Kreisel and Peter Hirschfeld for useful discussions.
This work was mainly supported by EPSRC (EP/I004475/1) and Oxford Centre for Applied Superconductivity.
Part of this work was supported by HFML-RU/FOM and LNCMI-CNRS, members of the European Magnetic Field Laboratory
(EMFL) and by EPSRC (UK) via its membership to the EMFL (grant no. EP/N01085X/1).
 We also acknowledge financial support of the John Fell Fund of the Oxford University.
AIC acknowledges an EPSRC Career Acceleration Fellowship (EP/I004475/1).

\bibliography{FeCuSe_bib_dec2021}

%%%%%
\newpage

\section{Appendix}

%\newcommand{\blue}{\textcolor{blue}}
%\newcommand{\bdm}[1]{\mbox{\boldmath $#1$}}
%
%\renewcommand{\thefigure}{S\arabic{figure}} % changes FIG.~1 to FIG.~SM1
%\renewcommand{\thetable}{S\arabic{table}} % changes FIG.~1 to FIG.~SM1

%\newlength{\figwidth}
%\figwidth=0.48\textwidth

%\setcounter{figure}{0}

%%\newcommand{\fig}[3]
%{
%	\begin{figure}[!tb]
%		\vspace*{-0.1cm}
%		\[
%		\includegraphics[width=\figwidth]{#1}
%		\]
%		\vskip -0.2cm
%		\caption{\label{#2}
%			\small#3
%		}
%\end{figure}}
%%%%

\begin{figure*}[htbp]
	\centering
	\includegraphics[trim={0cm 0cm 0cm 0cm}, width=0.8\linewidth,clip=true]{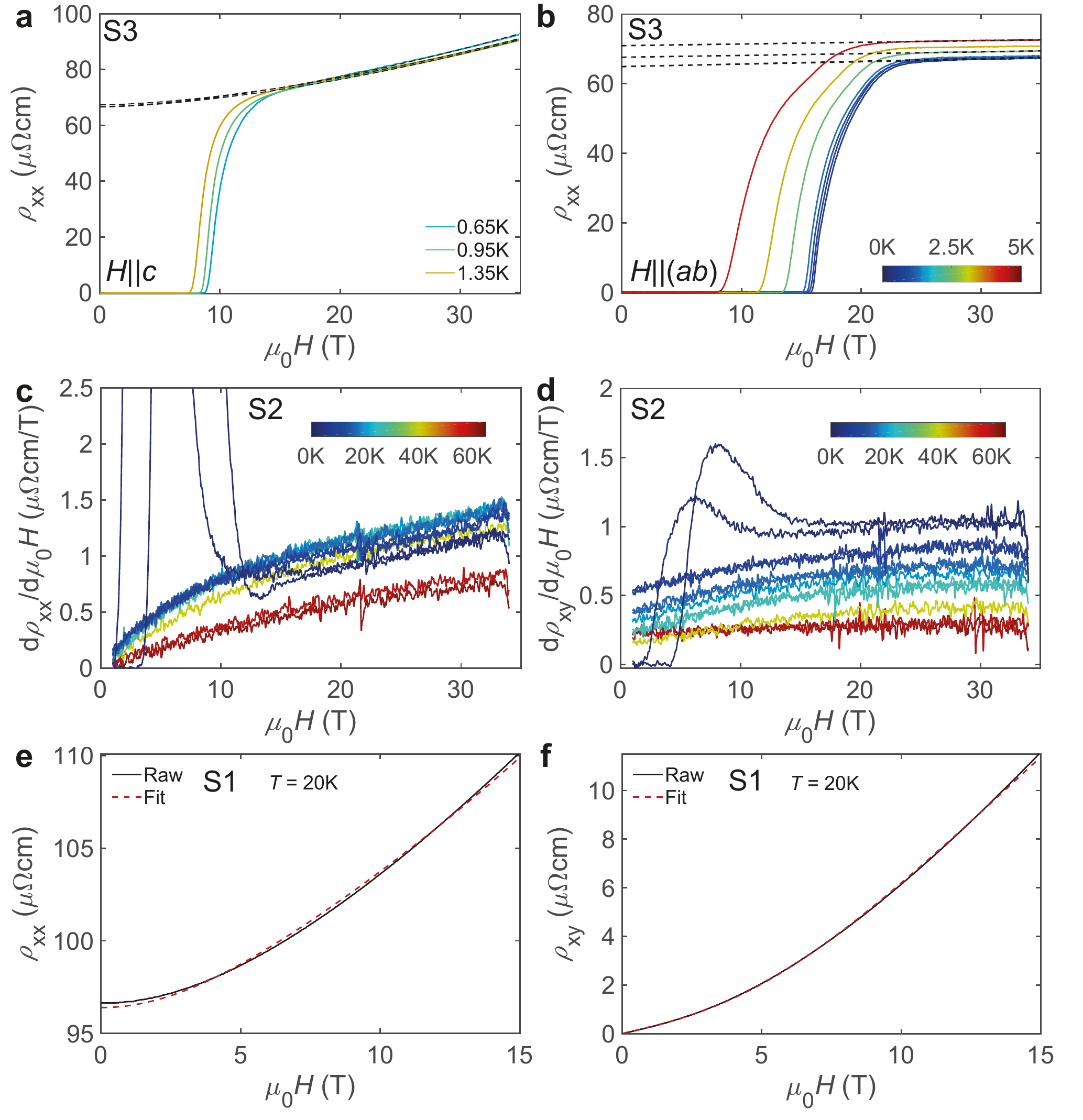}
	\caption{{\bf Magnetotransport behaviour of Fe$_{1-x}$Cu$_{x}$Se, $x = 0.0025$.}
	(a) Longitudinal resistivity, $\rho_{xx}$, with the magnetic field $H||c$ for sample S3.
Dashed lines are fits to a $H^{1.6}$ field dependence, in the normal state, to extrapolate the zero-field resistivity at low temperatures in Fig.~\ref{Fig_normal_state}c.
(b) Longitudinal resistivity, $\rho_{xx}$ with $H||$($ab$) for sample S3.
Dashed lines are fits to a linear dependence in magnetic field, in the normal state, to extrapolate the zero-field resistivity in magnetic field in Fig.~\ref{Fig_normal_state}c.
(c) The derivatives of $\rho_{xx}$ and (d) $\rho_{xy}$ in magnetic field of sample S2 (raw data in Fig.~\ref{Fig_x_2}(c) and (d)).
	An example of simultaneous fitting to a three-carrier model of the resistivity component
$\rho_{xx}$ in (e) and $\rho_{xy}$ in (f) (red dashed lines) measured at $T$ = 20~K.
}
	\label{Fig_SM_Mobility}
\end{figure*}

\end{document}